\overfullrule=0pt
\input harvmac

\input epsf

\def\a{\alpha}

\def\b{\beta}

\def\g{\gamma}

\def\d{\delta}

\def\e{\epsilon}

\def\l{\lambda}

\def\lh{\widehat\lambda}

\def\o{\omega}

\def\oh{\widehat\o}

\def\O{\Omega}

\def\t{\theta}

\def\G{\Gamma}

\def\L{\Lambda}

\def\N{\nabla}

\def\Pb{\overline\Pi}
\def\p{\partial}
\def\pb{\overline\partial}
\def\dh{\widehat d}
\def\Jb{\overline J}

\def\Tt{\widetilde T}

\Title{ \vbox{\baselineskip12pt \hbox{IFT-P.016/2008}}}
{\vbox{\centerline{ Yang-Mills Chern-Simons Corrections}
\smallskip
\centerline{From The Pure Spinor Superstring }}}

\bigskip
\centerline{Oscar A. Bedoya\foot{e-mail: abedoya@ift.unesp.br} \foot{
New address since August, 2008: Instituto de F\'{\i}sica,
Universidade de S\~ao Paulo, 05315-970, S\~ao Paulo, SP, Brasil.
e-mail:abedoya@fma.if.usp.br}}
\bigskip
\centerline{\it Instituto de F\'{\i}sica Te\'orica, State University
of S\~ao Paulo} \centerline{\it Rua Pamplona 145, 01405-900,
S\~ao Paulo, SP, Brasil. }

\bigskip

\noindent Nilpotency of the pure spinor BRST operator in a curved
background implies superspace equations of motion for the background.
By computing one-loop corrections to nilpotency for the heterotic
sigma model, the Yang-Mills Chern-Simons corrections to the
background are derived.

\Date{July 2008}


\newsec{Introduction} 
It is a well known fact that in order to couple
the Type I or Heterotic superstrings to a generic background, the gauge
groups must be $SO(32)$ or $E_8 \times E_8$ in order to have a theory
free of gauge and Lorentz anomalies. This condition is supplemented
with an $\a'$ correction to the 3-superform $H$, defined as the exterior
derivative of the Kalb-Ramond 2-superform B. The mechanism described is known as the Green-Schwarz mechanism
\ref\GreenSchwarz{M. Green and J. Schwarz, ``Anomaly Cancellations in
Supersymmetric $D=10$ Gauge Theory and Superstring Theory,'' Phys. Lett. B 149, 117 (1984)}
and the form of the corrections are of Yang-Mills and Lorentz Chern-Simons type, which is related to 
 the form of the counter-terms that cancel the anomalies. It is worth to note 
 that this mechanism for the cancellation of anomalies was discovered using
 the low energy limit of superstrings. However, Hull and Witten
 \ref\HullWitten{C.M Hull and E. Witten, Supersymmetric Sigma Models and the
Heterotic String, Phys. Let. B160 (1985) 398.}
 noted the necessity of the Chern-Simons modifications in order to cancel
 the sigma model for the Heterotic superstring. 

To describe superstrings in a generic background, one has at
disposal the Ramond-Neveu-Schwarz (RNS) formalism and the Green-Schwarz (GS)
formalism. However in the first, whose sigma model was the one used in \HullWitten, it is difficult to incorporate space-time
fermions, so some elements are lacking; while in the second one can only quantize in the light-cone gauge,
loosing the manifest symmetries. Nevertheless, there is one more description known as the Pure
Spinor (PS) formalism \ref\Berkovits{N. Berkovits, ``Super-Poincare
Covariant Quantization of the Superstring,'' JHEP 0004, 018 (2000) [arXiv:
hep-th/0001035].}, in which a superstring can be described in a generic
background \ref\BerkovitsHowe{N. Berkovits and P.S.
Howe, ``Ten-Dimensional Supergravity Constraints from the Pure Spinor 
Formalism for the Superstring'', Nucl. Phys. B 635, 75 (2002)
[arXiv:hep-th/0112160]} and does not suffer of those difficulties.
The quantization of the superstring in the PS formalism is performed through
a BRST charge $Q_{BRST} $, which is nilpotent because of the pure spinor
condition, to be defined later on. As shown in \BerkovitsHowe ,  
 the classical BRST invariance impose some constraints on the background
fields, in particular on the components of $H$; putting them on-shell.
Before  pure spinors were used to describe superstrings, integrability along
pure spinor lines allowed to find the super Yang-Mills and
supergravity equations of motion in ten dimensions \ref\Howe{P.S. Howe, ``Pure spinor
lines in superspace and ten-dimensional supersymmetric theories,''
Phys. Lett. B258: 141 (1991). }. Because
of its nature, the pure spinor sigma model is a proper description for
performing perturbative computations. Using this description it has been possible to compute the beta
functions for the Heterotic \ref\ChandiaVallilo{O. Chandia and B.C. Vallilo,
``Conformal Invariance of the Pure Spinor Superstring in a Curved
Background,'' JHEP 0404 (2004) 041 [arXiv:hep-th/0401226].} and Type II
Superstring \ref\BedoyaChandia{O.A. Bedoya and O. Chandia, ``One-loop
Conformal Invariance of the Type II Pure Spinor Superstring in a Curved
Background,'' JHEP 0701 (2007) 042, [arXiv:hep-th/0609161].}, showing that 
the classical BRST invariance implies in the conformal
invariance\foot{For further studies of the pure spinor formalism
in a curved background see \ref\Kluson{J. Kluson, ``Note about
Classical Dynamics of Pure Spinor String on $AdS_5 \times S^5$
Background,'' Eur. Phys. J. C50:1019 (2007)  [arXiv:hep-th/0603228]. J. Kluson ``Note About
Redefinition of BRST Operator for Pure Spinor String in General
Background,'' arXiv:0803.4390 [hep-th].}}.

Because in the PS formalism one can quantize in a Super-Poincare invariant
manner, one could attempt to compute $\a'$ corrections to the constrains in 
the background fields mentioned in the last paragraph. In particular, one can look for 
Chern-Simons type corrections to the $3$-superform $H$ as mentioned in the first
paragraph. This paper is concentrated in the Yang-Mills Chern-Simons
correction to $H$, which was also shown in \ref\AtickDharRatra{J. J. Atick,
A. Dhar and B. Ratra, ``Superspace formulation of ten-dimensional $N=1$
supergravity coupled to $N=1$ super Yang-Mills,'' Phys. Rev. D 33,
2824 (1986).} and \ref\HoweCS{P.S. Howe, ``Pure
Sinors, function superspaces and supergravity theories in ten and
eleven dimensions,'' Phys. Lett. B273: 90 (1991).} to imply the correct 
coupling of ${\cal N} =1$ supergravity to ${\cal N} =1$ super
Yang-Mills. Specifically, it will be computed corrections to the classical constraints on
$H$ by checking the nilpotency of the BRST charge at one-loop level. It will
be shown that it is a key aspect to add local counter-terms in the action to
preserve the BRST invariance at the quantum level. Those counter-terms 
amounts to redefinitions of the space-time metric and the spin
connection. The redefinition of the space-time metric was noted by Sen
\ref\Sen{A. Sen, ``Local Gauge And Lorentz Invariance Of The Heterotic
String Theory,'' Phys. Lett. B 166: 300, (1986).}. Furthermore Hull and
Townsend \ref\HullTownsend{C.M. Hull and P.K. Townsend, ``World-sheet
Supersymmetry and Anomaly Cancellation in the Heterotic String,'' Phys. Lett.
178: 187, (1986).}
showed that they were necessary to preserve the world sheet supersymmetry in
the heterotic string. Since the supervielbein $E_M{}^\a (Z)$ appears as one
of the superfields in the pure spinor sigma model, redefinitions of this
superfield are in accordance with redefinition of the space-time metric, and as will be
shown, they are important to check the BRST invariance at one-loop. 

The structure of this paper is as follows. In section 2 a brief 
introduction to the PS formalism is given. In section 3  the
results of \BerkovitsHowe\  and \ref\Chandia{O. Chandia, ``A note on the
Classical BRST Symmetry of the Pure Spinor String in a Curved Background,''
JHEP 0607 (2006) 019 [arXiv:hep-th/0604115]} concerning the nilpotency of
$Q_{BRST}$ and holomorphicity of the BRST current at the lowest order
in $\a'$ are reobtained, by performing a tree-level
computation.  In section 4 it is performed a one-loop computation to find
the Yang-Mills Chern-Simons correction to the $3$-superform $H$, explaining
the computations in a detailed way, as well as the counter-terms introduced.
In section 5 the work is concluded. In the appendix are included the results of 
the background field expansion used in the computation.

\newsec{Review of the Pure Spinor Formalism}
The action for the heterotic superstring in the pure spinor formalism
\Berkovits\ is given by 
\eqn\flataction{S = {1\over{2\pi \a'}}\int d^2 z ({1\over 2} \p X^m \pb X_m
+ p_\a \pb \t^\a + \bar b \p \bar c) + S_\l + S_{\Jb} , }
where the worldsheet variables $(X^m ,  \t^\a , p_\a)$, with $m = 0 {\ldots}
9$, $\a = 1{\ldots} 16$, describe the $N =1$ $D=10$ superspace. $p_\a$ is the
conjugate momentum to $\t^\a$. This formalism takes its name from the bosonic
spinor $\l^\a$, which is constrained to satisfy the pure spinor condition
$\l^\a (\g^m)_{\a\b} \l^\b =0$, where $\g^m$ are $16\times16$ symmetric
ten-dimensional gamma matrices. The pure spinor part of the action, denoted
by $S_\l$, is the action for a free
$\b$ $\g$ system, where the conjugate momentum to $\l^\a$ is denoted by
$\o_\a$. $S_{\Jb}$ denotes the action for the heterotic right-moving currents
and $(\bar b,\bar c)$ are the right moving Virasoro ghosts. For the purpose
of this paper, it is worth to note that the Lorentz currents $N^{ab} =
{1\over 2}\l \g^{ab}\o$ and ghost number current $J = \l^\a \o_\a $ satisfy 
\eqn\NN{N^{mn}(y) N^{pq}(z) \rightarrow \a'{{\eta^{p[n} N^{m]q}(z) - \eta^{q[n}
N^{m]p}(z)}\over{y-z}} -3\a'^2{{\eta^{m[q}\eta^{p]n}}\over{(y-z)^2}},}$$
J(y) J(z) \rightarrow -{{4}\over{(y-z)^2}}.$$ 
These currents have OPEs with the pure spinors
\eqn\curretsps{N^{mn}(y) \l^\a (z) \rightarrow {1\over 2} \a'(\g^{mn})^\a{} _\b
{{\l^\b (z)}\over{y-z}} , \,\,\,\,\,\, J(y) \l^\a (z) \rightarrow \a'{{\l^\a
(z)}\over{y-z}},}
while the right-moving currents satisfy
\eqn\JbJb{\Jb^I (y)\Jb^J (z) \rightarrow \a'{{f^{IJ}{}_K \Jb^K
(z)}\over{\bar y - \bar z}} + \a'^2 {{\d^{IJ}}\over{(\bar y - \bar z)^2}}.}

Physical states are defined as vertex operators in the cohomology of the
BRST charge $Q = \oint dz \l^\a d_\a$, where $d_\a$ are the worldsheet
variables corresponding to $N=1$ $D=10$ space-time supersymmetric derivatives.

\newsec{Lowest Order Constraints in $\a'$}
In this section are computed the constraints
coming from the nilpotency of the BRST charge and holomorphicity of the BRST
current at tree level.

The action which describes the Heterotic Superstring in a curved 
background can be obtained by adding the massless vertex operators
to the flat action and then covariantizing with respect to the 
$D=10$ $N =1$ super-reparameterization invariance \BerkovitsHowe\ . 
The action is as follows

\eqn\actioncurved{ S = {1\over{2\pi\a'}} \int d^2 z ({1\over 2}\Pi^a \Pb^b
\eta_{ab} + {1\over 2}\Pi^A \Pb^B B_{BA}+d_\a \Pb^\a + \Pi^A \Jb^I A_{AI} + d_\a \Jb^I
W_I ^\a }$$ \l^\a \o_\b \Jb^I U_{I\a}{}^\b + \l^\a \o_\b \Pb^C
\O_{C\a}{}^\b) + S_\l + S_{\Jb} + S_{\Phi}, $$
where $\Pi^A = \p Z^M E_{M}^A(Z)$, $\Pb^A = \pb Z^M E_M ^A (Z)$ and $E_M ^A (Z)$ 
is a supervielbein: $ G_{MN} (Z) = E_M ^a E_N ^b \eta_{ba} $. $Z^M$ denote
the coordinates for the $D=10$ $N=1$ superspace $(X^m ,\t^\mu)$ with $m =
0,{\ldots} ,9$ and $\mu = 1,{\ldots}, 16$. $S_\l$ and $S_{\Jb}$, as before, are the 
actions for $\l$ and $\Jb^I = {1\over {2}} {\cal K}_{\cal AB}^{I}
\bar\psi^{\cal A}\bar\psi^{\cal B}$ respectively, with ${\cal A, B }=
0,{\ldots} ,32$. $S_{\Phi}$ is the action for the dilaton coupling to the
worldsheet scalar curvature. The nilpotency of the BRST charge is guaranteed 
in a flat background because of the pure spinor condition. Nevertheless, when the superstring is coupled to the curved
background, the background fields must be constrained in order to maintain this
nilpotency \BerkovitsHowe\ \Chandia\ .  One can find these constrains by performing a tree level
computation. To set that, one perform a background field expansion
\ref\deBoerSkenderis{J. de Boer and K. Skenderis, ``Covariant Computation
of the Low Energy Effective Action of the Heterotic Superstring'', Nucl.
Phys. B 481, 129 (1996) [arXiv: hep-th/9608078].} by expliting every
worldsheet field into a classical and quantum part, where the classical
 part is assumed to satisfy the classical equation of motion and the quantum part
will allow to find propagators and form loops. Specifically, 
the following notation for the splitting will be used

\eqn\wsfsplitting{Z^M = X_0 ^M + Y^M , \,\,\, d_\a = d_{\a 0}+ \dh_\a,} $$
\l^\a = \l^\a _0 + \hat\l ^\a ,\,\,\, \o_\a = \o_{\a 0}+ \hat \o_\a , \,\,\,
\bar\psi ^{\cal A} = \bar\psi^{\cal A} _0 +\hat{\bar\psi}^{\cal A}. $$
So the expansion for the term ${1\over {2\pi \a'}} \int d^2 z {1\over 2}\p Z^M \pb Z^N G_{NM}$ 
in \actioncurved in second order of the quantum fiels is

\eqn\sndordereta{ {1\over{2\pi\a'}}\int d^2 z({1\over 2}\p Y^a \pb Y^b
\eta_{ab} - {1\over 2}\p Y^a Y^B \Pb^C \Tt_{CB}{}^a - {1\over 2}\pb Y^a Y^B
\Pi^C \Tt_{CB}{}^a  
 +{1\over 4}\p Y^B  Y^C \Pb^a \Tt_{CB}{}^a }$$ +{1\over 4}\pb Y^B Y^C \Pi^a
\Tt_{CB}{}^a + {1\over 2}Y^B Y^C \Pi^D
\Tt_{DC}{}^a \Pb^E \Tt_{EB}{}^a -{1\over 4}Y^B Y^C \Pi^{(a}\Pb^{D)} \Tt_{DCB}{}^a),$$
where $\Tt$ is the part of the torsion which only contains derivatives of
the vielbein: $\Tt_{MN}{}^A =  \p_{[M} E_{N]}{}^A$ and $\Tt_{DCB}{}^A =  - \Tt_{DC}{}^E \Tt_{EB}{}^A +(-)^{CD}\N_C
\Tt_{DB}{}^A$. Repeated bosonic indices in \sndordereta\ are assumed to be contracted
with the Minkowski metric. On the other hand, the expansion for ${1\over
{2\pi \a'}}\int d^2 z d_\a \pb
Z^M E_M{} ^\a$ is
\eqn\sndorderdbp{{1\over{2\pi \a'}} \int d^2 z (\dh_{\a} \pb Y^\a -\dh_\a Y^B 
\Pb^C \Tt_{CB}{}^\a +{1\over 2} (d_{\a 0} + \dh_\a ) \pb Y^B Y^C
\Tt_{CB}{}^\a } $$ -{1\over 2} (d_{\a 0}+ \dh_\a)Y^B \Pb^D Y^C (\p_C 
\Tt_{DB}{}^\a + \Tt_{CD}{}^E 
\Tt_{EB}{}^\a )+{1\over 2}\dh_\a \Pb^D Y^M Y^N \p_N E_{M}{}^B \Tt_{BD}{}^\a )$$
In the subsequent sections, the $0$ subindex will be dropped off. The expansions for the remaining terms in the expansion of the action
\actioncurved\ are written in the
appendix. From the first term in the last two expressions it can be read the
propagators 

\eqn\propagators{Y^a (x,\bar x)Y^b(z,\bar z) \to -\a' \eta^{ab}log|x-z|^2
,\,\,\, \dh_\a (x) Y^\b (z) \to {\a' \d_\a {}^\b\over{x-z}}.}

\subsec{Nilpotency at tree level}
The propagators \propagators\  allows to compute the conditions for the 
nilpotency of $Q_{BRST}$ perturbatively in $\a'$. In fact, one can easily
compute a tree level diagram using
the second propagator and the fifth term in \sndordereta\ expanding $e^{-S}$ 
in a series power, giving as a result

\eqn\fst{ \l^\a d_\a (w) \l^\b d_\b (z) = {1\over 2}\a' {1\over {w-z}}\l^\a \l^\b 
\Pi^c T_{\b\a}{}^c (z).}
Initially one is interested in computing the tree leve
diagrams coming from terms in the expansions with $\pb Y^A Y^B$, since
they will give rise to the same kind of pole as in \fst. So, the contributions
to the pole $(w-z)^{-1}$ will be

\eqn\polewz{{1\over 2} {\a' \over {w-z}}\l^\a \l^\b \Pi^c (T_{\b\a}{}^c
+H^c {}_{\b\a})(z) +{1\over 2}{\a' \over{w-z}}\l^\a \l^\b \Pi^\g H_{\g \b\a}
} $$+{\a' \over{w-z}} \l^\a \l^\b d_\g T_{\b\a}{}^\g (z) +{\a'
\over{w-z}}\l^\a \l^\b \l^\g \o_\d R_{\b\a\g}{}^\d (z). $$
In this notation, the Torsion superfield $T_{\b\a}{}^\g$ is given by
\eqn\Torsionfermionic{T_{\b\a}{}^\g = \Tt_{\b\a}{}^\g - \O_{\b\a}{}^\g -
\O_{\a\b}{}^\g ,}
while the curvature superfield is given by
\eqn\Curvature{ R_{\a\b\g}{}^\d = D_\a \O_{\b\g}{}^\d + D_\b \O_{\a\g}{}^\d +
\O_{\a\g}{}^\e \O_{\b\e}{}^\d + \O_{\b\g}{}^\e \O_{\a\e}{}^\d + \Tt_{\a\b}{}^E
\O_{E\g}{}^\d ,}
where $D_\a$ denotes the supersymmetric derivative. There are also other possible tree level contractions of $\l^\a d_\a (w) \l^\b
d_\b (z)$ with terms including $\p Y^A Y^B$ which will lead to 
\eqn\polebwbz{-{1\over 2} \a' {{\bar w -\bar z}\over {(w-z)^2}} \l^\a \l^\b
\Pb ^c (T_{\b\a}{}^c - H^c {}_{\a\b})(z) +{1\over 2}\a' {{\bar w - \bar
z}\over{(w-z)^2}}\l^\a \l^\b \Pb^\g H_{\g\a\b}(z)} $$ -\a' {{\bar w - \bar
z}\over {(w-z)^2}} \l^\a \l^\b \Jb^I F_{\a\b I}.$$
In this notation the field-strength superfield is given by
\eqn\fieldstrenth{F_{\a\b I} = D_\a A_{\b I} + D_\b A_{\a I} + f_I
{}^{JK}A_{\a J}A_{\b K} + \Tt_{\a\b}{}^C A_{C I}.}
To compute the tree-level diagrams that give rise to the above result, 
one need to compute the integral

\eqn\integralddpi{\int d^2 x {1\over {(w-x)(x-z)^2}} = -\int d^2 x \pb_x
{(\bar x -\bar w)\over {x-w}} {1\over{(x-z)^2}} = 2\pi {{\bar w- \bar z}\over
{(w-z)^2} }  }
From \polewz\ and \polebwbz\ it is deduced that the conditions for the nilpotency
of $Q_{BRST}$ at the lowest order in $\a'$ are
\eqn\nilpotencyconstraints{\l^\a \l^\b T_{\a\b}{}^C =0, \,\,\, \l^\a \l^\b
H_{C \a\b} =0, \,\,\, \l^\a\l^\b F_{\a\b I} = 0, \,\,\, \l^\a\l^\b\l^\g
\o_\d R_{\b\a\g}{}^\d =0.}
These are the same set of constraints found in \BerkovitsHowe\ and \Chandia\ .

\subsec{Holomorphicity at tree level}
To compute the conditions for holomorphicity of the BRST current $\pb j =
\pb (\l^\a d_\a ) =0$, one must know the expansion up to first order in $Y^\a$ of
the sigma model action. This expansion for the term ${1\over {2\pi \a'}}
\int d^2 z {1\over 2}\p Z^M \pb Z^N G_{NM}$ is 
\eqn\Gfstorder{{1\over{4\pi \a'}}\int d^2 [\Pi^a \pb Y^b \eta_{ab} + \Pb^a \p
Y^b\eta_{ab} + \Pi^b \Pb^D Y^C \Tt_{CD}{}^a \eta_{ab} + \Pi^D \Pb^a Y^C
\Tt_{CD}{}^b \eta_{ab} ]. }
The conditions for holomorphicity will appear as conditions for vanishing
to the independent couplings $\Pi^a \Pb^b$, $\Pi^\a \Pb^b$ and so on.
For example, forming a tree level diagram contracting $\pb d_\a$ in $\pb j$
with the third term in \Gfstorder\ , it is obtained ${1\over 2}\l^\a \Pi^b \Pb^C
\Tt_{C\a}{}^d \eta_{bd}$. Following this procedure with all the terms in the 
expansion written in the appendix up to order $Y$, it is found
\eqn\holtreelevel{{1\over 2} \l^\a [-\Pi^b \Pb^c (T_{\a b}{}^d \eta_{dc} +
T_{\a c}{}^d \eta_{bd}+ H_{cb\a}) + \Pi^\b \Pb^c (T_{\b\a b} - H_{\b\a b}) +
\Pi^b \Pb^\g (T_{\g\a b} + H_{\g\a b}) }$$ - \Pi^\b \Pb^\g H_{\g\b\a} 
-2d_\b \Pb^c T_{c\a}{}^\b -2d_\b \Pb^\g T_{\g\a}{}^\b+ 2 \Pi^b \Jb^I F_{b\a
I} +2 \Pi^\b \Jb^I F_{\b\a I} + 2 \l^\b \o_\g \Pb^d R_{d\a\b}{}^\g 
$$ $$- 2d_\b \Jb^I (D_\a W_I^\b-W_J ^\b A_{\a K}f_I{}^{JK}-U_{I\a}{}^\b ) + 2
\l^\b\o_\g \Jb^I (\O_{\a\d}{}^\g U_{I\b}{}^\d - \O_{\a\b}{}^\d U_{I\d}{}^\g
+ U_{J\b}{}^\g A_{\a K}f_I {}^{JK} $$ $$ 
- W_I^\d R_{\d\a\b}{}\g - D_\a U_{I\b}{}^\g )] = 0.$$ 
Since $\Pb^\a$ is related to $\Jb^I$ through $\Pb^\a = - \Jb^I W_I^\a$ by
using the equation of motion for the worldsheet field $d_\a$ in
\actioncurved\ , one arrives at the following set of constrints for
holomorphicity of the BRST current at the lowest order in $\a'$
\eqn\holomorphicityTL {T_{\a (bc)} = - H_{\a bc} = T_{\a\b}{}^c - H_{\a\b}{}^c = T_{c\a}{}^\b
 = 0, \,\, \l^\a \l^\b R_{d\a\b}{}^\g =0, \,\, F_{\a\b I} = - {1\over 2}W_I^\g
 H_{\g\a\b}, } $$ F_{\a bI} = - W^\g T_{\g\a b} , \,\,  \N_\a W_I^\b -
 T_{\a\b}{}^\g W_I^\g = U_{I\a}{}^\b , \,\, \l^\a \l^\b (\N_\a U_{I\a}{}^\g +
 R_{\a\g\b}{}^\d W_I ^\g) =0 .$$
This was the same set of constraints found in \BerkovitsHowe\ and \Chandia\ .
 
\newsec{Yang-Mills Chern-Simons Corrections}
In this section $\a'$ corrections to the nilpotency
constraints \nilpotencyconstraints\ will be computed.  In the first
subsection it is explained how to compute all of the
twenty possible contributions to the nilpotency of the BRST charge. In the second
subsection, it will be explained how, adding some counter-terms, one can find the
Yang-Mills Chern-Simons $3-$form.

\subsec{One-loop Corrections to the Constraints}
In the expansion for the $\Pi^A \Jb^I A_{A I}$ term,  the following
will play a role in the computation: $\Pi^A Y^B  _0 \Jb^I _2 (\p_B A_{AI}
+ \Tt_{BA}{}^C A_{C I})(x)$ and
$\p Y^A \Jb ^I _2 A_{AI}(y)$. Contracting them with $\l^\a d_\a (w)
\l^\b d_\b (z)$ one can form a 1-loop diagram
\eqn\picI{\eqalign{\eqalign{\epsfbox{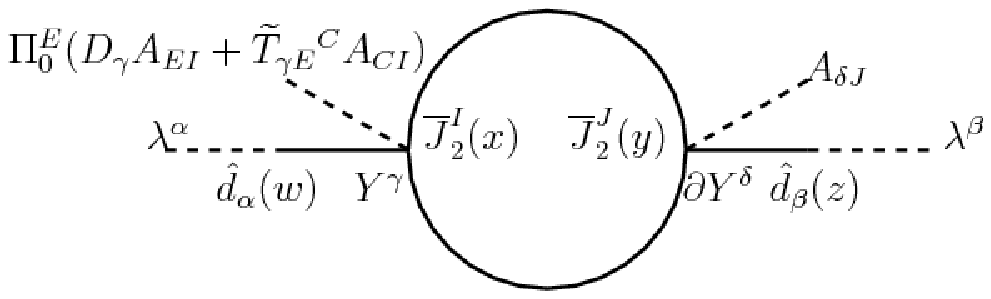} }}}
The dashed lines denote background fields while the continuous lines denote
the contractions using the propagators.
So one can compute how these terms contribute
to the nilpotency of $Q_{BRST}$. To determine the coefficient for this
diagram, note that there is an $1/2$ from the expansion of $exp[-S]$ and
there is a factor of $2$ coming from the possible ways to put the
superfields at $x$ or $y$. Denoting the integration over the world-sheet
fields by $\int [Dwsf]$, it is found 

\eqn\loopI{\l^\a d_\a (w) \l^\b d_\b (z)_{I}  = {1\over {(2\pi\a')^2}}\int [D
wsf]\int d^2 x d^2 y \l^\a \dh_\a (w) \l^\b \dh_\b (z) }$$\Pi^E _0 Y^\g (D_\g
A_{E I}+\Tt_{\g E}{}^F A_{F I})(x) \p Y^\d A_{\d J}(y) \Jb^I _2 (x) \Jb_2 ^J (y) $$ 
\eqn\intermezzoI{= {{\a'^2 }\over {(2 \pi)^2 }}  \l^\a \l^\b \Pi^C _0 A_{\a
I}(D_\b
A_{C I} + \Tt_{\b C}{}^D A_{D I})(z)
\int d^2 x d^2 y {1\over{(w-x)^2(z-y)}} {1\over{(\bar x - \bar y)^2}} } $$
 - {\a'^2 \over {(2\pi )^2}} \l^\a \l^\b \Pi^C _0 A_{\b I}(D_{\a} A_{C I}
 +\Tt_{\a C}{}^D A_{D I}) (z)\int
d^2 x d^2 y {1\over{(w-y)(z-x)^2}} {1\over{(\bar x - \bar y)^2}} ,$$
where $\Jb^I _2 (\bar x) \Jb^J _2 (\bar y) \to {(\a')^2\d^{IJ} \over{(\bar x - \bar y )^2}} $.
The second line in the last equation is obtained from minus the first by
interchanging $\a$ with $\b$ and $w$ with $z$. So, just one of the
integrals will be computed.
\eqn\integralI{\int d^2 x d^2 y {1\over {(w-x)^2(z-y)(\bar x - \bar y)^2}} =
\int d^2 x d^2 y {1\over {(w-x)^2(z-y)}}\pb_{\bar y} {1\over
{\bar x -\bar y}} } $$  = 2\pi \int d^2 x d^2 y {\d^2 (y-z) \over {(w-x)^2(\bar
x - \bar y)}} = 2\pi \int d^2 x {1\over{(w-x)^2 (\bar x -\bar z)}}
,  $$  
where in the second step an integratetion by parts has been performed with 
respect to $\bar y$.
In the last integral one can integrate by parts with respect to $x$ to obtain

\eqn\integralIR{\int d^2 x d^2 y {1\over {(w-x)^2(z-y)}} {1\over {(\bar x -
\bar y)^2}} = -{{(2\pi)^2} \over {w-z}}. }
Then a first contribution to the check of nilpotency will be
\eqn\QQI{\l^\a d_\a (w) \l^\b d_\b (z) _{I} = -2\a'^2  {{\l^\a
\l^\b}\over {w-z}} \Pi^C _0 A_{\b I}(D_\a A_{C I} +\Tt_{\a C}{}^D A_{D I})(z).}

A second contribution comes from contracting $\l^\a
d_\a (w) \l^\b d_\b (z)$ with $\p Y^\g \Jb^I _2 A_{\g I}(x) \times $ $\p Y^\d \Jb^J _2 A_{\d
J}(y)$ as shown in the diagram. 
\eqn\picII{\eqalign{\eqalign{\epsfbox{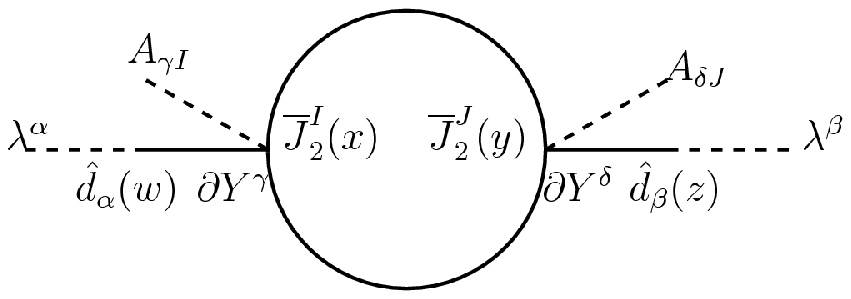} }}}
To determine the coefficient of this diagram, note that there is an
$1/2$ coming from the Taylor expansion of $exp(-S)$. So it is found

\eqn\loopII{\l^\a d_\a (w) \l^\b d_\b (z) _{II} = {\a'^2 \over 2 }{{\l^\a \l^\b
(z)}\over{(2\pi)^2}}\int d^2 x d^2 y [{{A_{\a I}(x)A_{\b
I}(y)}\over{(w-x)^2(z-y)^2}} - {{A_{\b I}(x)A_{\a
I}(y)}\over{(w-y)^2(z-x)^2}}] {1\over(\bar x -\bar y)^2}}
The second term in the integrand is obtained from minus the first by
interchanging $w$ with $z$ and $\a$ with $\b$. The integral left to
solve is
\eqn\intermezzoIIa{\G = \int d^2 x d^2 y {{A_{\a I}(x)A_{\b
I}(y)}\over{(w-x)^2(z-y)^2 (\bar x - \bar y)^2}} = -\int d^2 x d^2 y
{{\Pb^C \p_C A_{\a I}(x)A_{\b I}(y)}\over{(\bar y - \bar x)(w-x)^2(z-y)^2}}
}$$ + \int d^2 x d^2 y {{A_{\a I}(x)A_{\b I}(y) \p_x \d^2 (x-w)}\over{(\bar y -
\bar x)(z-y)^2}},$$
where it has been integrated by parts with respect to $\bar x$. The first and second
integral on the right hand side of \intermezzoIIa can be integrated by parts
with respect to $y$ and $x$ to obtain 
\eqn\intermezzoIIb{\G = 2\pi \int d^2 x d^2 y {{\Pb^C \p_C A_{\a I}(x)A_{\b
I}(y)\d^2 (y-x)}\over{(z-y)(w-x)^2}} - 2\pi \int d^2 x d^2 y {{\Pi^C \p_C
A_{\a I}(x)A_{\b I}(y)\d^2 (x-w)}\over{(\bar y -\bar x)(z-y)^2}}. }
Evaluating the superfields in $z$, using \integralddpi in the first integral and integrating by parts with
respect to $y$ in the second, one obtains
\eqn\intermezzoIIc{\G = -(2\pi)^2 {{\bar w -\bar z}\over{(w-z)^2}}\Pb^C \p_C
A_{\a I}A_{\b I}(z) - {{(2\pi)^2}\over{w-z}}\Pi^C \p_C A_{\a I}A_{\b I}(z).}
Then
\eqn\loopIIR{\l^\a d_\a (w) \l^\b d_\b (z) _{II} = -\a'^2 {{\bar w -\bar
z}\over{(w-z)^2}}\l^\a \l^\b \Pb^C \p_C
A_{\a I}A_{\b I}(z) - {\a'^2 \over{w-z}}\l^\a \l^\b \Pi^C \p_C A_{\a I}A_{\b
I}(z)} $$ +\a'^2  {{\bar w -\bar z}\over{(w-z)^2}}\pb \l^\a \l^\b A_{\a
I}A_{\b I} + {\a'^2 \over{w-z}}\p\l^\a \l^\b A_{\a I}A_{\b I}(z)$$ 

A third contribution to the nilpotency property comes from contractions of
$\Pi_0 ^A \Jb^I _2 A_{A I}$, twice  $\p Y^A \Jb_2
^I A_{AI}$ and $\l^\a d_\a (w) \l^\b d_\b (z)$ giving rise to the diagram
\eqn\picIII{\eqalign{\eqalign{\epsfbox{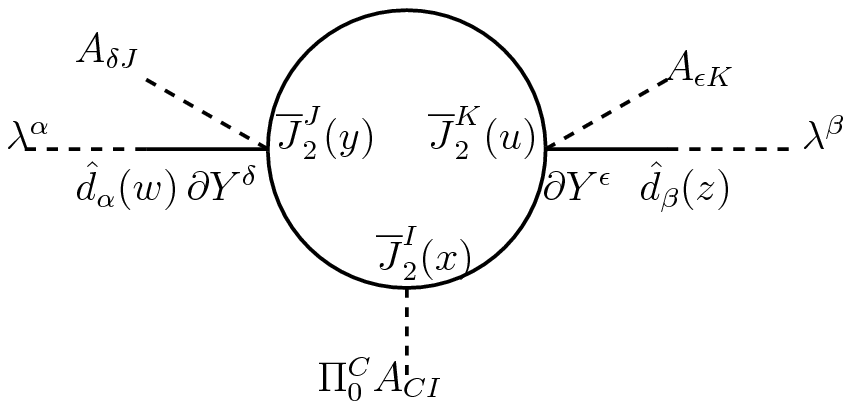} }}}
Since one is at order $S^3$ in the expansion of $e^{-S}$, there is an
${1\over{3!}}$ and also a factor of $3$ from the possible ways to put 
the superfields at $x$, $y$ and $u$, so there will be a $- 1/2$
coefficient in front:

\eqn\loopIII{\l^\a d_\a (w) \l^\b d_\b (z)_{III} = -{1\over{2(2\pi\a')^3}}\int
[Dwsf]\int d^2 x d^2 y d^2 u \l^\a \dh_\a (w) \l^\b \dh_\b (z) }$$\Pi^C _0 \Jb^I
_2 A_{C I}(x) \p Y^D \Jb_2 ^J A_{DJ}(y) \p Y^E \Jb_2 ^K A_{E K}(u). $$

\eqn\intermezzoIII{=-{1\over{2(2\pi)^3 \a'}} \l^\a \l^\b \Pi^C _0 A_{CI} A_{\g
J}A_{\d K}(z)\int d^2 x d^2 y d^2 u ({\d_\a {}^\g \d_\b {}^\d
\over{(w-y)^2(z-u)^2}} }$$ -{\d_\a {}^\d \d_\b {}^\g
\over{(w-u)^2(z-y)^2}})  \Jb_2 ^I (x) \Jb^J _2 (y) \Jb^K _{2} (u). $$
It is not hard to verify that
\eqn\JJJ{\Jb_2 ^I (x) \Jb^J _2 (y) \Jb^K _{2} (u) = {(\a')^3 f^{IJK} \over{(\bar x -
\bar y)(\bar y -\bar u)(\bar x - \bar u)}}+{\ldots} ,}
where by ${\ldots} $ is meant less singular poles which are not
important in this computation.
Then the type of integrals that must be computed are 
\eqn\integralIII{ \G_1 =\int d^2 x d^2 y d^2 u {1
\over{(w-y)^2(z-u)^2 (\bar x - \bar y)(\bar y -\bar u)(\bar x - \bar u)}}.}
The integral in $x$ gives

\eqn\integralx{\int d^2 x {1\over {(\bar x - \bar y)(\bar x - \bar u)}} =
\int d^2 x \partial_x ({{x-y}\over{\bar x - \bar y}}){1\over{\bar x - \bar u}} =-2\pi {{y-u}\over{\bar y -\bar u}},}
so \integralIII\ yields
\eqn\integralIIIa{\G_1 = -2\pi \int d^2 y d^2 u \p_y ({1\over w-y}) {y-u \over
(z-u)^2 (\bar y - \bar u )^2}.}
Integrating by parts in $y$, $\bar y$ and then in $u$ it is found $\G_1 = (2\pi)^3
/(w-z)$. In this way \loopII\ gives
\eqn\loopIIIR{\l^\a d_\a (w) \l^\b d_\b (z) _{III}= - (\a')^2 {\l^\a \l^\b \over
w-z }f^{IJK} \Pi^C _0 A_{CI} A_{\a J} A_{\b K} (z).}

Note that a fourth loop could be formed with ${1\over 4}\pb Y^\a  Y^\b \Pi^c
(T_{\b\a}{}^c+H^c{}_{\b\a})$, $\dh_\a \Jb^I _2 W_I ^\a$ and $\p Y^\a \Jb^I _2
A_{\a I}$ as shown in the diagram below.
\eqn\picIV{\eqalign{\eqalign{\epsfbox{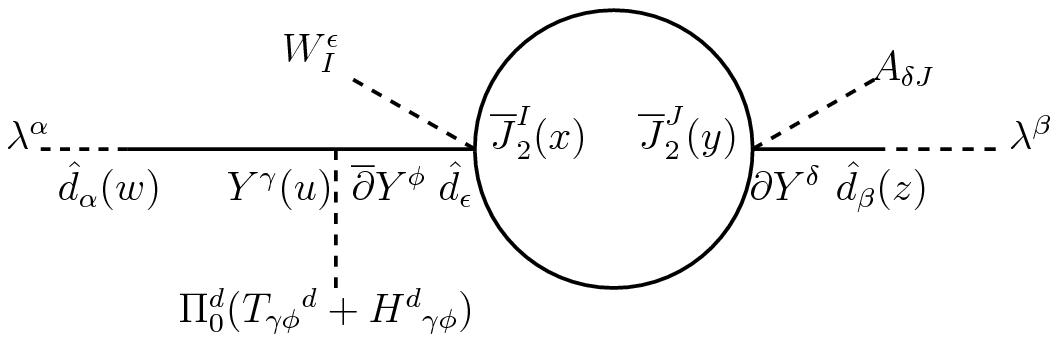} }}}
In this case, one is also at the order $S^3$, so there is an ${1\over {3!}}$
which is cancelled by the symmetry factor responsible for the localization 
of the superfields, either at $x$, $y$ or $u$. The $1\over 4$ coming from the
coefficient of the term with $\Pi^c$ is cancelled by a symmetry factor of the
possible ways of contraction:

\eqn\loopIV{\l^\a d_\a (w) \l^\b d_\b (z)_ {IV} = -{{\a'^2}\over{(2\pi)^2} }
\l^\a \l^\b \Pi^c (T_{\d\a}{}^c + H^c {}_{\d\a})W^\d _I A_{\b I}(z)
\times } $$ \int d^2 x
d^2 y d^2 u {{\d^2 (x-w) }\over{(z-u)^2(y-x)(\bar y - \bar u)^2}} $$
Integrating $x$ one has to solve
\eqn\intermezzoIVa{\int d^2 y d^2 u {1\over{(z-u)^2(y-w)(\bar y - \bar
u)^2}} = -2\pi \int d^2 y d^2 u {{\d^2 (y-w) }\over {(\bar u - \bar
y)(z-u)^2}} = -{{(2\pi )^2}\over{w-z}}.}

Then 
\eqn\loopIVR{\l^\a d_\a (w) \l^\b d_\b (z)_{IV} = {{\a'^2}\over{w-z}}\l^\a \l^\b
\Pi^c (T_{\a\d}{}^c + H^c {}_{\a\d})W^\d _I A_{\b I}(z)}

Considering the same last diagram but with the vertex ${1\over 4} \Pi^\g
H_{\g\b\a}$ instead of ${1\over 4} \Pi^c(T_{\b\a}{}^c + H_{\b\a}{}^c)$,
gives a fifth contribution to the coupling to $\Pi^\g$
\eqn\loopVR{\l^\a d_\a (w) \l^\b d_\b (z)_{V} = {{\a'^2}\over{w-z}}\l^\a \l^\b
\Pi^\g H_{\g\a\d}W^\d _I A_{\b I}(z)}

A sixth contribution can be formed with ${1\over 4}\Pi^c \pb Y^A Y^B (\Tt_{BA}{}^c + H^c {}_{BA})$
and twice $\p Y^A \Jb^I _2 A_{AI}$:
\eqn\picV{\eqalign{\eqalign{\epsfbox{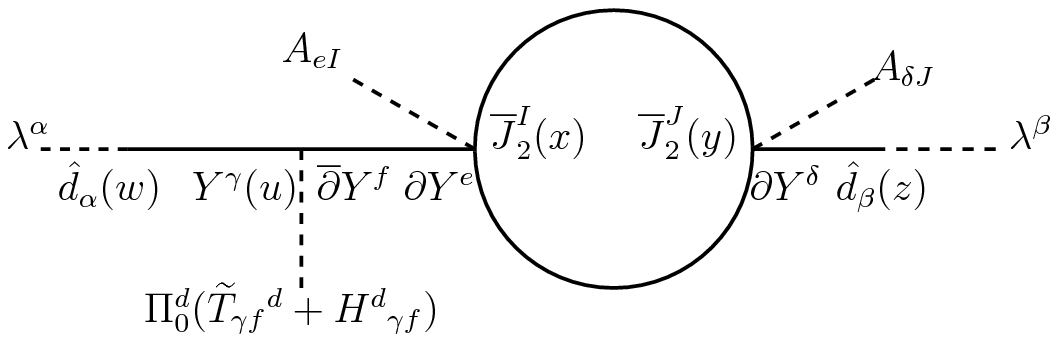} }}}
There are $8$ possible ways of making the
contractions, a $3$ factor from the possible ways to put the superfields at
$x$, $y$ or $u$, an $1/3!$ because one is at $S^3$ in the expansion, and the
factor of $1/4$ of the $\Pi^c$ term gives a one coefficient:

\eqn\loopVI{\l^\a d_\a (w) \l^\b d_\b (z)_{VI} = -{{\a'^2}\over{(2\pi
)^2}}\l^\a \l^\b \Pi^c(\Tt_{d\a}{}^c + H^c {}_{d\a})A_{d I}A_{\b I}(z) \times } $$ \int d^2 x
d^2 y d^2 u {{\d^2 (x-w) }\over{(y-x)(z-u)^2}} {1\over{(\bar  y
-\bar u)^2}.}  $$
The integral is the same as in \loopIV , so the answer is 
\eqn\loopVIR{\l^\a d_\a (w) \l^\b d_\b (z)_{VI} = {{\a'^2}\over{w-z}}\l^\a \l^\b
\Pi^c (\Tt_{d\a}{}^c + H^c {}_{d\a})A_{d I}A_{\b I}(z).}

In the same way, the last diagram but with the vertex ${1\over 4}\Pi^\g
H_{\g BA}$ instead of ${1\over 4} \Pi^c (T_{BA}{}^c + H_{BA}{}^c)$ leads to
a seventh contribution 
\eqn\loopVIIR{\l^\a d_\a (w) \l^\b d_\b (z)_{VII} = {{\a'^2}\over{w-z}}\l^\a \l^\b
\Pi^\g H_{\g d\a}A_{d I}A_{\b I}(z).}

An eight contribution can be formed with $-{1\over 2} \pb Y^a Y^\b
\Pi^C \Tt_{C\b}{}^a$ and twice $\p Y^A \Jb^I _2 A_{AI}$:
\eqn\picVI{\eqalign{\eqalign{\epsfbox{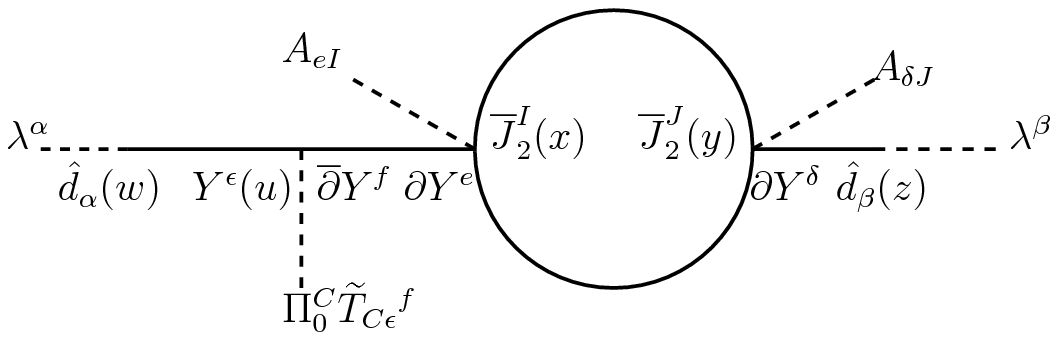} }}}
There are $4$ possible
ways of making the contractions, a $3$ factor from the possible ways to put
the superfields at $x$, $y$ or $u$, an $1\over 3!$ because one is at $S^3$
order in the expansion and a factor of $1/2$ of the $\Pi^a$ coefficient,
giving at the end a $1$ coefficient:

\eqn\loopVIII{\l^\a d_\a (w) \l^\b d_\b (z)_{VIII} = -{{\a'^2}\over{(2\pi )^3}}
\l^\a \l^\b \Pi^C \Tt_{C\a}{}^d A_{\b I} A_{d I}(z) \times } $$ \int d^2x d^2 y d^2 u
{{-2\pi \d^2 (u-x)}\over{(w-x)(z-y)^2}} {1\over{(\bar u - \bar y)^2
}}.$$
Integrating in $u$, the integral one has to solve is

\eqn\intermezzoVIIIa{\int d^2 x d^2 y {1\over{(w-x)(z-y)^2(\bar x - \bar y)^2}} = 2\pi \int  d^2 x d^2 y 
{{\d^2 (x-w) }\over{(z-y)^2 (\bar y - \bar x)}}  =  {{(2\pi)^2 }\over{w-z}} ,}
then
\eqn\loopVIII{\l^\a d_\a (w) \l^\b d_\b (z)_{VIII} = {{\a'^2}\over{w-z}}\l^\a
\l^\b \Pi^C \Tt_{C\a}{}^d A_{\b I} A_{d I}(z).}

Let's consider the couplings to $\Pb^A$.

A diagram like \picIV\ can be formed with ${1\over 4} \Pb^c \p Y^A Y^B
 (\Tt_{BA}{}^c - H^c {}_{BA})$, $\p Y^A \Jb^I _2 A_{AI}$ and $\dh_\a \Jb^I _2
W_I ^\a$. There are $4$ possible
ways of making the contractions, a $6$ factor from the possible ways to put
the superfields at $x$, $y$ or $u$, an $1\over 3!$ because one is at $S^3$
order in the expansion and a factor of $1/4$ of the $\Pb^c$ coefficient,
giving at the end a $1$ coefficient to this ninth contribution:

\eqn\loopIX{\l^\a d_\a (w) \l^\b d_\b (z)_{IX} =  {{\a'^2}\over{(2\pi
)^3}} \l^\a \l^\b \Pb^c (T_{\d\a}{}^c - H^c {}_{\d\a}) W^\d _I A_{\b I}(z)
\times } $$\int
d^2 x d^2 y d^2 u {1\over{(w-x)^2(z-u)^2 (y-x)(\bar y - \bar u)^2}}$$
Integrating $\bar y$ by parts, one is left to solve the integral
\eqn\intermezzoIXa{\int d^2x d^2 y d^2 u {{\d^2 (y-x)}\over{(w-x)^2 (z-u)^2
(\bar u - \bar y)}} = 2\pi \int d^2 x {1\over{(w-x)(z-x)^2}}.}
The right hand side in the last equation is the same as \integralddpi , so
\eqn\loopIXR{\l^\a d_\a (w) \l^\b d_\b (z)_{IX} = - \a'^2 {{\bar w -
\bar z}\over{(w-z)^2}} \l^\a \l^\b \Pb^c (T_{\d\a}{}^c - H^c
{}_{\d\a})W^\d _I A_{\b I}(z).}

In the same way, considering vertex $-{1\over 4} \Pb^\g H_{\g BA}$ instead of
$-{1\over 4} \Pb^c (\Tt_{BA}{}^c - H_{BA}{}^c )$ leads to the tenth contribution

\eqn\loopX{\l^\a d_\a (w) \l^\b d_\b (z)_{X} = \a'^2 {{\bar w -
\bar z}\over{(w-z)^2}} \l^\a \l^\b \Pb^\g H_{\g\d\a}W^\d _I A_{\b I}(z)}

An eleventh contribution comes from a diagram like \picV\ which can be formed with ${1\over 4} \Pb^c \p Y^A Y^B
 (\Tt_{BA}{}^c - H^c {}_{BA})$ and twice $\p Y^A \p \Jb^I _2 A_{AI}$ . There are
$8$ possible ways of making the contractions, a $3$ factor from the possible ways to put
the superfields at $x$, $y$ or $u$, an $1\over 3!$ because one is at $S^3$
order in the expansion and a factor of $1/4$ of the $\Pb^c$ coefficient,
giving at the end a $+$ coefficient: 

\eqn\loopXI{\l^\a d_\a (w) \l^\b d_\b (z)_{XI} = {{\a'^2}\over{(2\pi)^3}}
\l^\a \l^\b\Pb^c (\Tt_{d\a}{}^c - H^c {}_{d\a})A_{dI}A_{\b I}(z) \times } $$\int d^2 x d^2
y d^2 u {1\over{(w-x)^2(z-u)^2(y-x)(\bar u - \bar y)}}.$$
The last integral is the same as the integral in \loopIX , so the result is
\eqn\loopXIR{\l^\a d_\a (w) \l^\b d_\b (z)_{XI} = -\a'^2 {{\bar w -\bar
z}\over{(w-z)^2}}\l^\a \l^\b \Pb^c (\Tt_{d\a}{}^c - H^c {}_{d\a})A_{d I}A_{\b
I}(z).}

In the same way, a twelfth contribution comes from considering the vertex
$-{1\over 4}\Pb^\g H_{\g BA}$ instead of the vertex ${1\over 4} \Pb^c
(\Tt_{BA}{}^c - H_{BA}{}^c)$, leading to 
\eqn\loopXII{\l^\a d_\a (w) \l^\b d_\b (z)_{XII} = \a'^2 {{\bar w - \bar
z}\over{(w-z)^2}}\l^\a \l^\b \Pb^\g H_{\g d\a}A_{dI}A_{\b I}(z).}

Another diagram like \picVI\ can be formed with $-{1\over 2} \p Y^a Y^\b
\Pb^C \Tt_{C\b}{}^a$, $\p Y^a \Jb^I _2 A_{a I}$ and $\p Y^\a \Jb^I _2 A_{\a
I}$, giving rise to a thirteenth contribution

\eqn\loopXIII{\l^\a d_\a (w) \l^\b d_\b (z)_{XIII} = -\a'^2 {{\bar w -\bar
z}\over{(w-z)^2}} \l^\a \l^\b \Pb^C \Tt_{C\a}{}^d A_{d I}A_{\b I}(z).}

A fourteenth contribution and the last for the couplings to $\Pb^A$ can be formed with $-\dh_\a Y^B \Pb^C
\Tt_{CB}{}^\a$ and twice $\p Y^A \Jb A_{AI}$:
\eqn\picVII{\eqalign{\eqalign{\epsfbox{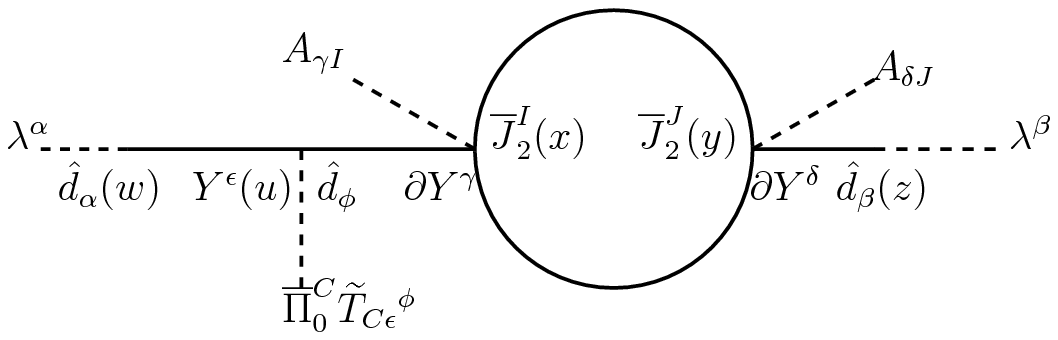} }}}
giving as result
\eqn\loopXIVR{\l^\a d_\a (w) \l^\b d_\b (z) _{XIV} = 2\a'{{\bar w - \bar
z}\over{(w-z)^2}} \l^\a \l^\b \Pb^C A_{\b I}\Tt_{C\a}{}^\g A_{\g I}}

Let's consider the couplings to $\Jb^I _0$

A fifteenth contribution to the nilpotency will come from a diagram formed with ${1\over 2} \p Y^A Y^B \Jb^I _0
(\p_{[B}A_{A]I} + \Tt_{BA}{}^C A_{C I})$, $\dh_\a \Jb^I _2 W_I ^\a$ and $\p
Y^\a \Jb^I _2 A_{\a I}$:
\eqn\picVIII{\eqalign{\eqalign{\epsfbox{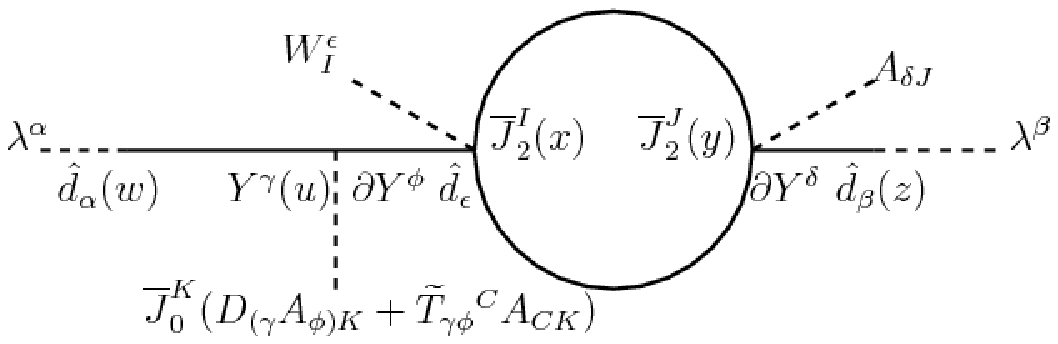} }}}
There are $4$
possible ways of making the contractions, a $6$ factor from the
possible ways to put the superfields at $x$, $y$ or $u$, an $1\over
3!$ because one is at the $S^3$ order in the expansion and a factor of $1/2$ of the $\Jb_0 ^I$ coefficient, giving at the end a 2 factor:
\eqn\loopXV{\l^\a d_\a (w) \l^\b d_\b (z)_{XV} =
{{2\a'^2}\over{(2\pi)^3}}\l^\a \l^\b \Jb^I _0 (D_{(\g}A_{\a )I} +
\Tt_{\g\a}{}^C A_{C I})W_J ^\g A_{\b J} (z) \times } $$ \int d^2 x d^2 y d^2 u
{1\over{(w-x)^2(z-u)^2(y-x)(\bar u - \bar y)^2}}.$$
The last integral is again the same as in \loopIX , so the result is 
\eqn\loopXVR{\l^\a d_\a (w) \l^\b d_\b (z)_{XV} = -2\a'^2 {{\bar w -\bar
z}\over{(w-z)^2}}\l^\a \l^\b \Jb^I _0 (D_{(\g}A_{\a )I} + \Tt_{\g\a}{}^C A_{C I})W_J ^\g A_{\b J} (z).}

A sixteenth contribution can be formed with ${1\over 2} \p Y^A Y^B
\Jb^I _0 (\p_{[B}A_{A]I} + \Tt_{BA}{}^C A_{C I})$ and twice $\p Y^A \Jb^I _2
A_{A I}$:
\eqn\picIX{\eqalign{\eqalign{\epsfbox{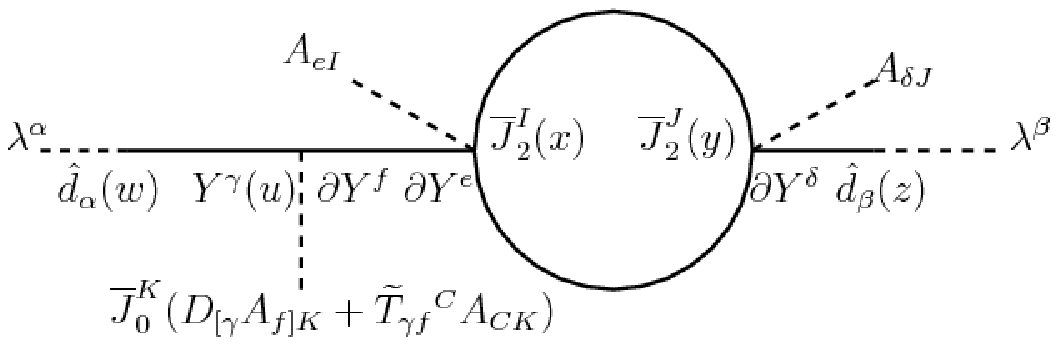} }}}
There are $8$ possible ways of making the contractions, a $3$ factor
from the possible ways to put the superfields at $x$, $y$ or $u$, an
$1\over 3!$ because one is at the $S^3$ order in the expansion and a factor of $1/2$ of the $\Jb^I _0$ coefficient, giving at the end a $2$ coefficient:
\eqn\loopXVI{\l^\a d_\a (w) \l^\b d_\b (z)_{XVI} =
2{{\a'^2}\over{(2\pi)^3}}\l^\a \l^\b \Jb_0 ^I (\p_{[c}A_{\a]I} + \Tt_{c\a}{}^D
A_{D I})A_{c J}A_{\b J} (z)\times }$$ \int d^2 x d^2 y d^2 u {1\over{(w-x)^2
(z-u)^2 (y-x)(\bar y - \bar u)^2}},$$
which contains the same integral as before, so the result is  
\eqn\loopXVIR{\l^\a d_\a (w) \l^\b d_\b (z)_{XVI} = -2 \a'^2 {{\bar w - \bar
z}\over{(w-z)^2}}\l^\a \l^\b \Jb^I _0 (\p_{[c}A_{\a]I} + \Tt_{c\a}{}^D A_{D I})A_{c J}A_{\b J} (z).}

Finally, let's consider the couplings to $d_\a$.

A seventeenth contribution can be formed with ${1\over2} d_\a \pb Y^\b Y^\g
\Tt_{\g\b}{}^\a$, 
$\dh_\a \Jb^I _2  W_I ^\a$ and $\p Y^\a \Jb^I _2 A_{\a I}$:
\eqn\picX{\eqalign{\eqalign{\epsfbox{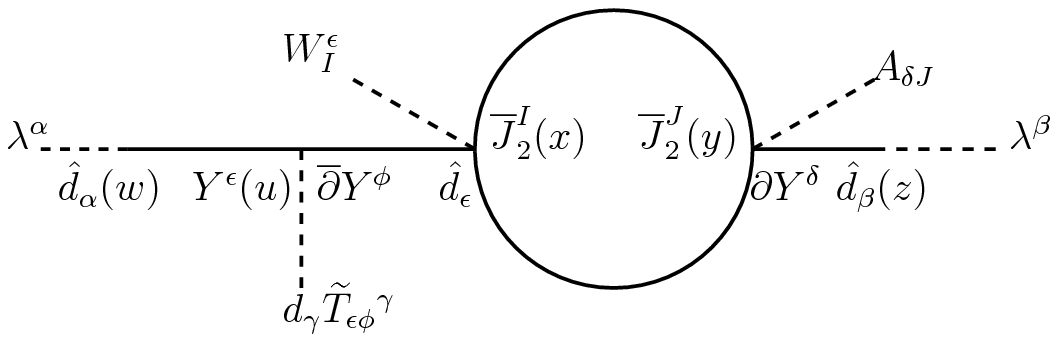} }}}
There are $4$ possible
ways of making the contractions, a $6$ factor from the possible ways to put the
superfields at $x$, $y$ or $u$, an $1\over 3!$ because one is at the $S^3$ order in the
expansion and a factor of $1/2$ of the $d_\a$ coefficient, giving at the end a $2$
coefficient:

\eqn\loopXVII{\l^\a d_\a (w) \l^\b d_\b (z)_{XVII} =
-2{{\a'^2}\over{(2\pi)^2}}\l^\a \l^\b d_\g \Tt_{\d\a}{}^\g
 W^\d _I A_{\b I} (z) \times } $$\int d^2 x d^2 y d^2 u {{\d^2
 (x-w)}\over{(z-u)^2(y-x)(\bar y - \bar u)^2}} $$
Integrating $x$, the integral that is left to solve is
\eqn\intermezzoXVI{\int d^2 y d^2 u {1\over{(z-u)^2(y-w)(\bar y - \bar
u)^2}} = -2\pi \int d^2 y d^2 u {{\d^2 (y-w)}\over{(\bar u - \bar
y)(z-u)^2}} = -{{(2\pi )^2}\over{w-z}},}
So, 
\eqn\loopXVIIR{\l^\a d_\a (w) \l^\b d_\b (z)_{XVII} = {{2\a'^2}\over{w-z}}
\l^\a \l^\b d_\g \Tt_{\d\a}{}^\g W_I ^\d A_{\b I}(z).}

An eighteenth contribution can be formed with ${1\over2} d_\a \pb Y^B Y^C \Tt_{CB}{}^\a$ and twice
$\p Y^A \Jb^I _2 A_{A I}$:
\eqn\picXI{\eqalign{\eqalign{\epsfbox{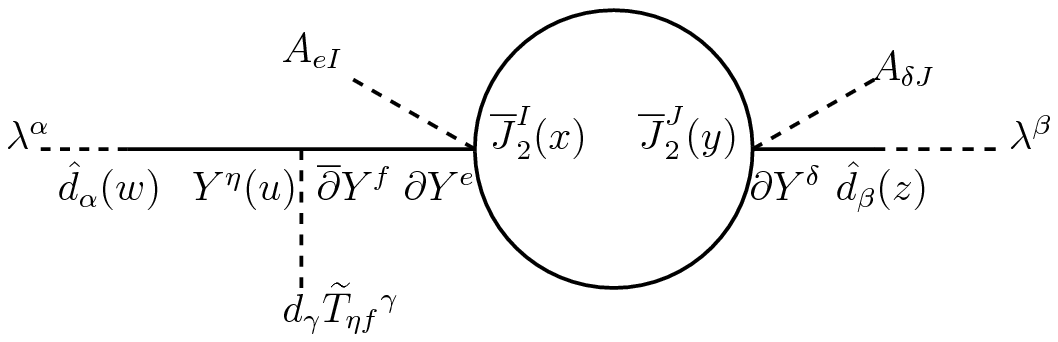} }}}
There are $8$ possible ways of making the contractions, a $3$ factor
from the possible ways to put the superfields at $x$, $y$ and $u$, an
$1\over 3!$ because one is 
at the $S^3$ order in the expansion and a factor of $1/2$ of the $d_\a$ coefficient , giving a $2$ coefficient:

\eqn\loopXVIII{\l^\a d_\a (w) \l^\b d_\b (z)_{XVIII} = 2{{\a'^2}\over{(2\pi
)^2}} \l^\a \l^\b d_\g \Tt_{c\a}{}^\g A_{cI}A_{\b I}(z)\times }$$ \int d^2 x
d^2 y d^2 u {{\d^2 (x-w)}\over{(z-u)^2 (y-x)(\bar y - \bar u)^2}}.$$
This integral is the same as in \intermezzoXVI , so the result is 
\eqn\loopXVIIIR{\l^\a d_\a (w) \l^\b d_\b (z)_{XVIII} = -{{2\a'^2}\over{w-z}}\l^\a \l^\b d_\g 
\Tt_{c\a}{}^\g A_{cI} A_{\b I}(z).}

Because of the pure spinor condition, the action is invariant under $\d
\o_\a = (\L_b \g^b \l)_\a$, so $U_{I\a}{}^\b = U_I \d_\a {}^\b + {1\over
4}U_{I cd}(\g^{cd})_\a {}^\b$. It can be formed a nineteenth one-loop diagram by
contracting $J \Jb^I_2 U_I (x)$ with $\p Y^\a \Jb^I _2 A_{\a I}$:
\eqn\picXII{\eqalign{\eqalign{\epsfbox{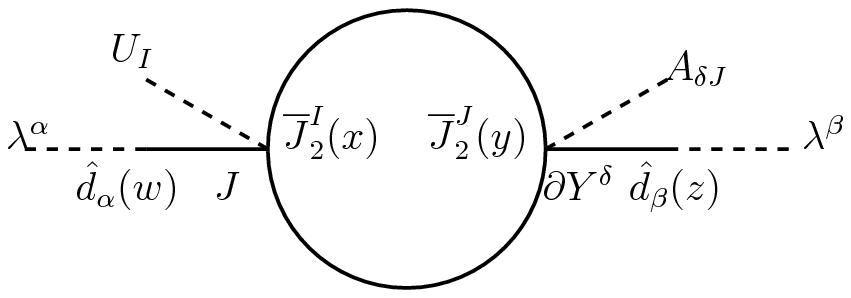} }}}
giving the contribution 
\eqn\loopXIXR{\l^\a d_\a (w) \l^\b d_\b (z) _{XIX} = -2{{\a'^2}\over{w-z}}
\l^\a \l^\b d_\g \d_\a {}^\g A_{\b I} U_I}

Similarly, a diagram like \picXII\ can be formed contracting ${1\over
2}N^{ab}\Jb^I_2
U_{I ab}(x)$ with $\p Y^\a \Jb^I_2 A_{\a I}$, giving as contribution
\eqn\loopXXR{\l^\a d_\a (w) \l^\b d_\b (z) _{XX} = -{1\over 2}{{\a'^2}\over{w-z}}
\l^\a \l^\b d_\g (\g^{ef})_\a {}^\g U_{I ef}A_{\b I}}

Now, the results will be summarized by adding up the twenty one-loop contributions to the tree level
constraints. Each independent worldsheet coupling will receive corrections,
as indicated below:

Corrections to the the coupling to $\Pi^c$ 
\eqn\treeplusonelooppic{{1\over 2} {\a' \over {w-z}}\l^\a \l^\b \Pi^c [(T_{\b\a}{}^c
+H^c {}_{\b\a}) -4\a' A_{\b I}(D_\a A_{c I} +\Tt_{\a c}{}^D A_{D I}) +
2\a'A_{\b I} \p_c A_{\a I}}$$ -2\a'f^{IJK} A_{cI} A_{\a J} A_{\b K} +2\a' (T_{\a\d}{}^c +
H^c {}_{\a\d})W^\d _I A_{\b I} +2\a' (T_{d\a}{}^c + T_{c\a}{}^e \eta_{ed}+
H^c {}_{d\a})A_{d I}A_{\b I}](z).$$

Corrections to the coupling to $\Pb^c$ 
\eqn\treeplusonelooppb{-{1\over 2} \a' {{\bar w -\bar z}\over {(w-z)^2}} \l^\a \l^\b
\Pb ^c [(T_{\b\a}{}^c - H^c {}_{\a\b}) - 2\a' A_{\b I} \p_c A_{\a I}+2\a'(T_{\d\a}{}^c - H^c
{}_{\d\a})W^\d _I A_{\b I} }$$ + 2\a'(T_{d\a}{}^c  + T_{c\a}{}^e \eta_{ed}-
H^c {}_{d\a})A_{d I}A_{\b I} -4\a'A_{\b I}\Tt_{c\a}{}^\g A_{\g I}](z). $$

Corrections to the coupling to $\Pi^\g$ 
\eqn\treeplusonelooppig{{1\over 2}{\a' \over{w-z}}\l^\a \l^\b \Pi^\g [H_{\g
\b\a} -4\a' A_{\b I}(D_\a A_{\g I} + \Tt_{\a \g}{}^D A_{D I}) -2\a'A_{\b
I}D_\g A_{\a I} }$$ - 2\a'f^{IJK}  A_{\g I} A_{\a J} A_{\b K}+ 2\a' H_{\g\a\d}W^\d _I
A_{\b I} +2\a' (T_{\g\a d} - H_{\g \a d})A_{d I}A_{\b I}](z).$$

Corrections to the coupling to $\Pb^\g$ 
\eqn\treeplusonelooppbg{{1\over 2}\a' {{\bar w - \bar
z}\over{(w-z)^2}}\l^\a \l^\b \Pb^\g [H_{\g\a\b} -2\a'A_{\b I }D_\g A_{\a I}+ 2\a'H_{\g\d\a}W^\d _I
A_{\b I}-2\a'(H_{\g \a }{}^d + T_{\g\a}{}^d )A_{dI}A_{\b I} }$$ + 4\a'A_{\b
I}\Tt_{\g\a}{}^\d A_{\d I}](z).$$

Corrections to the coupling to $d_\g$ 
\eqn\treeplusoneloopdg{{\a' \over{w-z}} \l^\a \l^\b d_\g [T_{\b\a}{}^\g +
2\a'\Tt_{\d\a}{}^\g W_I^\d A_{\b I} -2 \a'\Tt_{c\a}{}^\g A_{cI}A_{\b I}
-2\a'U_{I\a}{}^\g A_{\b I}].}

Corrections to the coupling to $\Jb^I _0$ 
\eqn\treeplusoneloopJb{-\a' {{\bar w - \bar
z}\over {(w-z)^2}} \l^\a \l^\b \Jb^I [F_{\a\b I} + 2\a' (D_{(\g}A_{\a )I} +
\Tt_{\g\a}{}^C A_{C I})W_J ^\g A_{\b J}}  $$+ 2\a' (\p_{[c}A_{\a]I} + \Tt_{c\a}{}^D
A_{D I})A_{c J}A_{\b J} ](z).$$

\subsec{Addition of Counter-terms}
Let's now concentrate in finding the Yang-Mills Chern-Simons $3-$form by
adding appropiate counter-terms. Keeping in mind the lowest order in $\a'$ holomorphicity constraints $T_{\a bc} + T_{\a cb} = 0 =
H_{\a bc}$; the conditions for nilpotency at one loop look like

From the coupling to $\Pi^c$
\eqn\nilpotencyonelooppi{\l^\a \l^\b [(T_{\b\a}{}^c
+H^c {}_{\b\a}) -4\a' A_{\b I}(D_\a A_{c I} +\Tt_{\a c}{}^D A_{D I}) +
2\a'A_{\b I} \p_c A_{\a I}}$$ -2\a'f^{IJK} A_{cI} A_{\a J} A_{\b K} +2\a' (T_{\a\d}{}^c +
H^c {}_{\a\d})W^\d _I A_{\b I} ](z) =0.$$

From the coupling to $\Pb^c$
\eqn\nilpotencyonelooppb{\l^\a \l^\b [(T_{\b\a}{}^c - H^c {}_{\a\b}) - 2\a'
A_{\b I} \p_c A_{\a I} +2\a'(T_{\d\a}{}^c - H^c
{}_{\d\a})W^\d _I A_{\b I} -4\a'A_{\b I}\Tt_{c\a}{}^\g A_{\g I}](z) =0 }
Adding \nilpotencyonelooppi\ and \nilpotencyonelooppb\ gives the condition
\eqn\Tcorrection{\l^\a \l^\b [T_{\b\a}{}^c -2\a' A_{\b I}(D_\a A_{c I} +\Tt_{\a c}{}^D A_{D I})
-\a'f^{IJK} A_{cI} A_{\a J} A_{\b K} +2\a' T_{\a\d}{}^c W^\d _I A_{\b
I} } $$-2\a'A_{\b I}\Tt_{c\a}{}^\g A_{\g I}]=0  .$$
Substracting \nilpotencyonelooppi\ and \nilpotencyonelooppb\ gives the condition
\eqn\Hcorrection{\l^\a \l^\b [H^c {}_{\b\a}
-2\a' A_{\b I}(D_{[\a} A_{c ]I} +\Tt_{\a c}{}^D A_{D I})-\a'f^{IJK} A_{cI}
A_{\a J} A_{\b K} +2\a' H_{\a\d}{}^c W^\d _I A_{\b I} } $$ + 2\a'A_{\b I}\Tt_{c\a}{}^\g A_{\g I}] =0. $$

Now, suppose that a counter-term of the form ${K_1\over{2\pi}}\int d^2 z \p Z^M \pb
Z^N A_{NI}A_{MI}$ is added to the action, where $K_1$ is a constant to be determined.
This amounts to redefine the space-time metric $G_{MN} \rightarrow G_{MN}
+ 2\a' K_1 A_{MI} A_{NI}$. The expansion of this counter-term will
contain the terms
\eqn\countertermexpn{S_C = {K_1\over {2\pi}} \int d^2 x [\p Y^A \pb Y^B A_{BI}
A_{AI} + \p Y^A \Pb^B A_{BI}Y^C (\p_C A_{AI}+{1\over 2}\Tt_{CA}{}^D
A_{DI} ) + } $$ \p Y^A \Pb^B Y^C (\p_C A_{BI} +\Tt_{CB}{}^D A_{DI})A_{AI} + \Pi^A \pb Y^B
A_{BI}Y^C (\p_C A_{AI}+\Tt_{CA}{}^D A_{DI})+ $$ $$\Pi^A \pb Y^B Y^C (\p_C
A_{BI} + {1\over 2}\Tt_{CB}{}^D A_{DI})A_{AI} ] $$
which can be used to compute tree level diagrams contracting with $\l^\a
\dh_\a (w) \l^\b \dh_\b (z)$. However this diagrams will
contribute to the order $\a'^2$, entering at the same foot as the one-loop
diagrams. The result of these tree level diagram is
\eqn\fstcc{-\a'^2 K_1 {{\bar w-\bar z}\over{(w-z)^2}}\l^\a \l^\b \Pb^C [ A_{C
I}( D_{(\a}A_{\b) I} +\Tt_{\a\b}{}^D A_{DI}) -2 A_{\b I} (D_\a A_{C I}
+ \Tt_{\a D}{}^D A_{DI})](z)} $$ \a'^2 K_1 {{\l^\a \l^\b
}\over{w-z}} \Pi^C [A_{C I} (D_{(\a} A_{\b ) I}
+\Tt_{\a\b}{}^D A_{DI})-2A_{\b I} (D_\a A_{C I} + \Tt_{\a C}{}^D A_{DI})](z)$$ $$
+2\a'^2 K_1 {{\bar w - \bar z}\over{(w-z)^2 }} \pb \l^\a \l^\b A_{\a I}A_{\b
I}(z) + 2\a'^2{{K_1 }\over{w-z}} \p\l^\a \l^\b A_{\a I} A_{\b I}(z)$$
Then, \nilpotencyonelooppi\ and \nilpotencyonelooppb\ will be modified
respectively to 
\eqn\nilpotencyonelooppic{\l^\a \l^\b [(T_{\b\a}{}^c
+H^c {}_{\b\a}) -4\a' A_{\b I}(D_\a A_{c I} +\Tt_{\a c}{}^D A_{D I}) +
2\a'A_{\b I} \p_c A_{\a I}}$$ -2\a'f^{IJK} A_{cI} A_{\a J} A_{\b K} +2\a' (T_{\a\d}{}^c +
H^c {}_{\a\d})W^\d _I A_{\b I} + 2\a'K_1 A_{c I}(D_{(\a} A_{\b )I}
+\Tt_{\a\b}{}^D A_{DI}) $$ $$-4\a'K_1
A_{\b I} (D_\a A_{cI} + \Tt_{\a c}{}^D A_{DI})](z) =0.$$
\eqn\nilpotencyonelooppbc {\l^\a \l^\b [(T_{\b\a}{}^c - H^c {}_{\a\b}) - 2\a'
A_{\b I} \p_c A_{\a I} +2\a'(T_{\d\a}{}^c - H^c
{}_{\d\a})W^\d _I A_{\b I} } $$+ 2\a'K_1 A_{c I}(D_{(\a} A_{\b )I} +\Tt_{\a\b}{}^D
A_{DI}) -4\a'K_1
A_{\b I} (D_\a A_{cI} + \Tt_{\a c}{}^D A_{DI}) -4\a'A_{\b I}\Tt_{c\a}{}^\g A_{\g I}](z) =0 $$
One can add \nilpotencyonelooppic\ with \nilpotencyonelooppbc\
to obtain 
\eqn\Tresult{\l^\a \l^\b [T_{\b\a}{}^c -2\a'A_{\b I}(D_\a A_{c I} +\Tt_{\a
c}{}^D A_{D I}) -\a'f^{IJK} A_{cI} A_{\a J} A_{\b K} +2\a' T_{\a\d}{}^c W^\d
_I A_{\b I} }$$+ 2\a'K_1 A_{c I}(D_{(\a} A_{\b )I} +\Tt_{\a\b}{}^D A_{DI})
-4\a'K_1 A_{\b I} (D_\a A_{cI} + \Tt_{\a c}{}^D A_{DI}) -2\a'A_{\b I}\Tt_{c\a}{}^\g A_{\g I}] =0 .$$
If $K_1 = -1/2$ and using the constraint $\l^\a \l^\b F_{\a\b I} = 0$
one arrives at 
\eqn\Tresultb{\l^\a \l^\b [T_{\b\a}{}^c + 2\a' T_{\a\d}{}^c W^\d
_I A_{\b I} -2\a'A_{\b I}\Tt_{c\a}{}^\g A_{\g I}] =0.}
Furthermore, forming a three-level diagram with $\dh_\a Y^\b \Pb^C \Tt_{C
\b}{}^\a $ and $\p Y^\a \pb Y^\b A_{\b I} A_{\a I}$ in \countertermexpn\ ,
with precisely this value for $K_1$ one can cancel the term proportional to 
$A_{\b I }\Tt_{c \a}{}^\g A_{\g I}$ in \Tresultb\ and \Hcorrection\ . Also,
with this value for $K_1$, the 
counter-terms in the last line of \fstcc\ will cancel the contributions
proportional to $\p \l^\a$ and $\pb \l^\a$ in \loopIIR\ .

Note that it can be added a second counter-term of the form ${{K_2}\over{2\pi}}\int
d^2 z d_\a \pb Z^M A_{MI}W_I ^\a $. This amounts to redefining the
supervielben $E_{M}{}^\a \rightarrow E_{M}{}^\a +\a' K_2 A_{MI}W_I ^\a$.
After expanding this counter-term, one can form a tree-level diagrams contracting 
it with ${1\over 4}\pb Y^\g Y^\d \Pi^c
(T_{\d\g}{}^c + H_{\d\g}{}^c)$:
\eqn\picXIII{\eqalign{\eqalign{\epsfbox{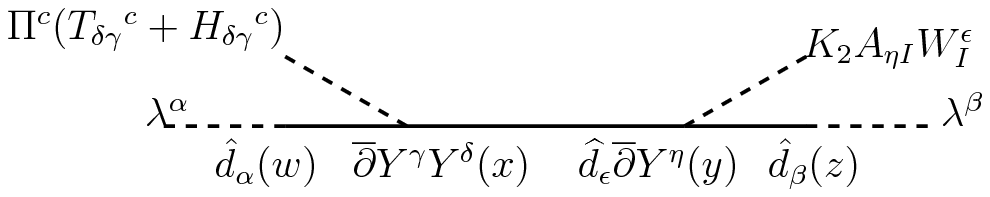} }}}
giving a contribution to the nilpotency
\eqn\KIIa{\a'^2 K_2 {{\l^\a \l^\b }\over{w-z}} \Pi^c (T_{\a\g}{}^c +
H_{\a\g}{}^c )W_I ^\g A_{\b I}(z), }
while contractions with ${1\over 4}\p Y^\g Y^\d \Pb^c
(T_{\d\g}{}^c - H_{\d\g}{}^c)$ will form the diagram
\eqn\picXIV{\eqalign{\eqalign{\epsfbox{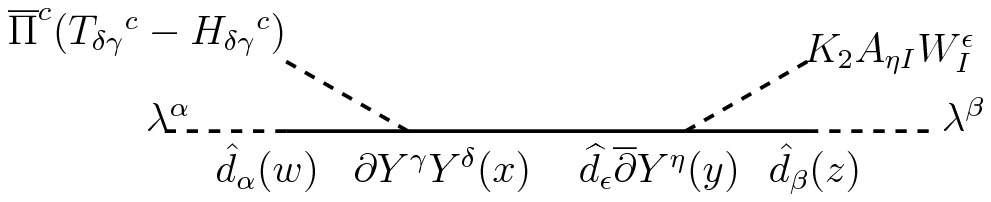} }}}
which gives the contribution
\eqn\KIIb{- \a'^2 K_2 {{\bar w - \bar z}\over{(w-z)^2}} \l^\a \l^\b \Pb^c
(T_{\a\g}{}^c - H_{\a\g}{}^c) W_I ^\g A_{\b I}.}
It can be easily checked that for $K_2 = -1$, adding \KIIa\ and \KIIb\ to
\nilpotencyonelooppi\ and \nilpotencyonelooppb\ respectively; then $\l^\a \l^\b
T_{\a\b}{}^c$ will not receive $\a'$ corrections, i.e. this second counter-term 
cancels the $\a'$ correction in \Tresultb; while the corrections for
$H_{\a\b}{}^c$ are
\eqn\HcorrectionIMZI{\l^\a \l^\b [H^c {}_{\b\a}
-2\a' A_{\b I}(D_{[\a} A_{c ]I} +\Tt_{\a c}{}^D A_{D I})-\a'f^{IJK} A_{cI}
A_{\a J} A_{\b K} ] =0. }

Now, the couplings to $\Pi^\g$ also receive corrections from the two
counter-terms just introduced. Some of these corrections come from the
coupling to $\Pi^C$ in \fstcc\ when $C$ is $\g$. Another correction comes
from the tree-level diagram
\eqn\picXV{\eqalign{\eqalign{\epsfbox{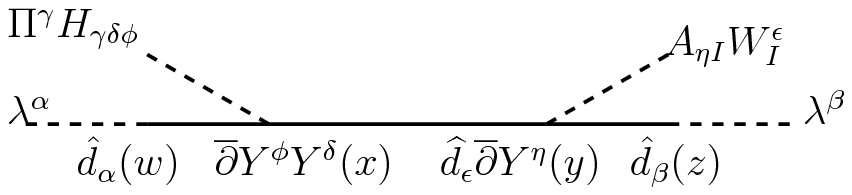} }}}
Adding those corrections and using the holomorphicity constraint $F_{\a\b I}
= -{1\over 2}W^\g _I H_{\g\a\b}$, it can be checked that the $\a'$ corrections to
the coupling to $\Pi^\g$ are
\eqn\HcorrectiongIMZI{\l^\a \l^\b [H_{\g\b\a}
-2\a' A_{\b I}(D_{(\a} A_{\g ) I} +\Tt_{\a \g}{}^D A_{D I})-\a'f^{IJK} A_{\g I}
A_{\a J} A_{\b K} ] =0. }

Let's now identify the Chern-Simons form. It can be used the lowest order constraints in $\a'$ coming from
nilpotency condition $\l^\a \l^\b F_{\a\b I} =0$ to write
\HcorrectionIMZI\ in the desired form. Since $\l^\a \l^\b = \l^\b \l^\a $
\eqn\HcorrectionIMZII{\l^\a \l^\b [H^c {}_{\a\b} -\a'Tr
A_{[\a}(D_{\b}A_{c]}+{1\over 2}\Tt_{\b c ]}{}^D A_{D]})
-2\a'f^{IJK}A_{cI} A_{\a J} A_{\b K} ](z) =0}
Since $2f^{IJK}A_{cI} A_{\a J} A_{\b K} = {2\over 3}Tr A_{[c}A_\a A_{\b]}$ then 
\eqn\HcorrectionIMZII{\l^\a \l^\b [H^c {}_{\a\b} -\a'Tr(A_{[\a}D_{\b}A_{c]} + {2\over
3} A_{[c} A_\a A_{\b ]} +{1\over 2}A_{[\a}\Tt_{\b c]}{}^D A_{D})](z) =0,}
which is the desired form.
Similarly, \HcorrectiongIMZI\ can be written as 
\eqn\HcorrectionIMZII{\l^\a \l^\b [H_{\a\b\g} -\a'Tr(A_{(\a}D_{\b}A_{\g)} + {2\over
3} A_{(\g} A_\a A_{\b )} +{1\over 2}A_{(\a}\Tt_{\b \g)}{}^D A_{D} )](z) =0.}

Adding a further third counter-term $-{1\over{2\pi}}\int d^2 z \l^\a \o_\b \pb
Z^M A_{M I}U_{I \a}{}^\b $, which amounts to redefine $\O_{M\a}{}^\b
\rightarrow \O_{M\a}{}^\b -\a'A_{M I}U_{I\a}{}^\b$;
and thanks also to the other two counter-terms added, can verify that neither 
$\l^\a \l^\b T_{\a\b}{}^\g = 0$ nor $\l^\a \l^\b F_{
\a\b I} =0$ will receive $\a'$ corrections.

\newsec{Conclusions}
The process of finding the Yang-Mills Chern-Simons 
correction to the 3-superform $H$ from a string
computation has been successful, in agreement with the studies of super Yang-Mills
and supergravity couplings \GreenSchwarz\ , \AtickDharRatra\ and \HoweCS. It is
interesting to note that to preserve worldsheet symmetries, some
redefinitions of the superfields are in order. Particularly, it was found that
for the pure spinor sigma model, both $E_M{}^a$ and $E_M{}^\a$ should be
redefined. The redefinition of the second one could not be found using the
other descriptions for the superstring. 

The procedure used in this paper is suitable for computing the Lorentz Chern-Simons 
$3$-superform in a pretty similar way, because there is a direct
analogy of the terms $\p Z^M \Jb^I A_{MI}$ and $\l^\a \o_\b \pb Z^M
\O_{M\a}{}^\b$ in the action.  In that case, diagrams formed by
contractions of terms with three quantum fields would contribute. Work in this direction is very interesting,
because a solution recently \ref\LechnerTonin{K. Lechner and
M. Tonin, ``Superspace formulations of ten-dimensional supergravity,''
arXiv: 0802.3869[hep-th]} has been claimed for the old debate about the
inclusion of the Lorentz Chern-Simons-form in $N =1$ $D = 10$ supergravity
and the $\a'$ corrections to the supergravity constraints. See 
\ref\BellucciGates{S. Bellucci and S. J. Gates, Jr., ``D =10 , N =1
Superspace Supergravity And The Lorentz Chern Simons Form,'' Phys.
Lett.B208: 456, (1988).}, \ref\BellucciDepireuxGates{S. Bellucci, D.A.
Depireux and S.J. Gates Jr., ``Consistent And Universal Inclusion Of The
Lorentz Chern-Simons Form In D=10, N=1 Supergravity Theories,'' Phys. Lett.
B238:315, (1990)}, \ref\GatexKissMerrell{S.J. Gates, Jr., A. Kiss and W.
Merrell, ``Dynamical equations from a first-order perturbative superspace
formulation of 10D, N=1 string-corrected supergravity (I),'' JHEP 0412:047,
(2004), [arXiv:hep-th/0409104]} for the {\it perturbative} approach and
\ref\AuriaFreRaciti{R. D'Auria, P. Fr{\`e}, M. Raciti and F. Riva, ``Anomaly
Free Supergravity In D=10.1. The Bianchi Identities And The Bosonic
Lagrangian,'' Int. J. Mod. Phys. A3:953, (1988)}, \ref\RacitiTivaZanon{M.
Raciti, F. Riva and D. Zanon, ``Perturbativa Approach to D=10 Superspace
Supergravity With a Lorentz Chern-Simons Form, '' Phys. Lett. B227:118,
(1989).} , \ref\Bonora\Pasti{L. Bonora, P. Pasti and M. Tonin, ``Superspace
Formulation Of 10-D Sugra+Sym Theory A La Green-Schwarz, '' Phys. Lett.
B188: 335, (1987); L. Bonora, M. Bregola, K. Lechner, P. Pasti and M. Tonin,
``Anomaly Free Supergravity And  Superyang-Mills Theories In Ten-Dimensions,
'' Nucl.Phys.B296:877, (1988).}, \ref\Bonoraetal{L. Bonora et al, ``Some
remarks on the supersymmetrization of the Lorentz Chern-Simons in D=10 N=1
supergravity theories,'' Phys. Lett. B277:306, (1992).} for the {\it
non-perturbative} approach. The pure spinor formalism was also used at the
cohomological level in \ref\ChandiaTonin{O. Chandia and M. Tonin, ``BRST anomaly and superspace
constraints of the pure spinor heterotic string in a curved background,'' JHEP
0709:016, 2007m arXiv: 0707.0654[hep-th].} to
study the BRST anomaly. It would be very interesting to perform a one-loop
computation to find the Lorentz Chern-Simons form, and relate the pure
spinor supergravity constraints with those in \LechnerTonin\ .

\newsec{Acknowledgements:} I would like to thank Osvaldo Chandia and
Vladimir Pershin for discussions and especially N. Berkovits for
valuable discussions and suggestions. The work of O.B is supported by CAPES, grant
33015015001P7.

\newsec{Appendix}

\subsec{Background Field Expansions}
From the expansion of the term ${1\over{2\pi\a'}}\int d^2 z{1\over 2}\p Z^M \pb Z^N B_{NM}$
\eqn\expansionB{{1\over {2\pi \a'}}\int d^2 z [{1\over 2}\Pi^B \Pb^A Y^C H_{CAB} + {1\over 4}Y^A \p Y^B \Pb^C
H_{CBA}-{1\over 4}Y^A
\pb Y^B \Pi^C H_{CBA} + {1\over 4}Y^A Y^B \Pi^C \Pb^D H_{DCBA}],}
where $H_{ABC} = (-)^{a(b+n)+(c+p)(a+b)}3 E_C ^P E_B ^N E_A ^M
\p_{[M}B_{NP]}$,
\eqn\defHMNP{\p_{[M}B_{NP]} = {1\over 3 }(\p_M B_{NP} + (-)^{m(n+p)}\p_N B_{PM} +
(-)^{p(m+n)}\p_p B_{MN})} and $H_{DCBA} = (-)^{B(C+D)}\N_B H_{DCA} - (-)^{BC}T_{DB}{}^E H_{ECA} +
(-)^{D(B+C)}T_{CB}{}^E H_{EDA}.$

From the expantion of $ {1\over{2\pi\a'}}\int d^2 z \p Z^M \Jb^I A_{M I}$
\eqn\expansionA{{1\over {2\pi\a'}}\int d^2 z [(\Jb^I_0 + \Jb^I_1 + \Jb^I_2)(\p Y^A A_{AI} + \Pi^A
 Y^B (\p_B A_{AI} + \Tt_{BA}{}^C A_{CI})+\Pi^A A_{AI}  }$$  + {1\over 2} \p Y^A  Y^B
(\p_{[B}A_{A] I} + \Tt_{BA}{}^C A_{CI}) +{1\over 2}  Y^A Y^B \Pi^C
\Tt_{CB}{}^D (\p_{D}A_{A I} + \Tt_{DA}{}^E A_{E I}) $$ $$ -{{(-)^{BC}}\over
2} Y^A Y^B \Pi^C \p_B (\p_{C}A_{A I} +\Tt_{CA}{}^D
A_{DI})  $$

From the expansion of ${1\over{2\pi\a'}}\int d^2 z d_\a \pb Z^M E_{M}^\a$
\eqn\expansionE{{1\over{2\pi\a'}}\int d^2 z [(d_{\a 0} + \dh_\a)(\pb Y^\a +
\Pb^B Y^C \Tt_{CB}{}^\a)],}
where the terms quadratic in $Y$ were written in \sndorderdbp .

From the expansion of ${1\over{2\pi\a'}}\int d^2 z d_\a \Jb^I W_I ^\a$
\eqn\expansionW{{1\over {2\pi\a'}}\int d^2 z [(d_{\a 0} + \dh_{\a})(\Jb^I_0
+ \Jb^I _1 + \Jb^I_2 )({1\over 2} Y^B
Y^C \p_C \p_B W_I ^\a  +Y^C \p_C W_I ^\a+ W_I ^\a ).}

From the expansion of ${1\over{2\pi\a'}}\int d^2 z \l^\a \o_\b \Pb^C \O_{C\a}{}^\b$
\eqn\expansionO{{1\over {2\pi\a'}}\int d^2 z [ (\lh^\a \o_\b +
\l^\a \oh_\b + \lh^\a \oh_\b) ({1\over 2}\pb Y^D Y^C (\p_{[C} \O_{D]\a}{}^\b +
\Tt_{CD}{}^E \O_{E\a}{}^\b) +  \Pb^C \O_{C\a}{}^\b }$$+{1\over 2}Y^C Y^D
\Pb^E \Tt_{ED}{}^F
(\p_{F} \O_{C \a}{}^\b + \Tt_{FC}{}^G \O_{G\a}{}^\b ) + \pb Y^C
\O_{C\a}{}^\b + \Pb^C Y^D (\p_D \O_{C\a}{}^\b + \Tt_{DC}{}^E \O_{E\a}{}^\b)$$
$$ -{1\over 2}(-)^{DE} Y^C Y^D \Pb^E \p_D (\p_{E}\O_{C\a}{}^\b + \Tt_{EC}{}^F
\O_{F\a}{}^\b))].$$ 

From the expansion of ${1\over{2\pi\a'}}\int d^2 z \l^\a \o_\b \Jb^I U_{I\a}{}^\b$
\eqn\expansionU{{1\over {2\pi\a'}}\int d^2 z [(\l^\a \o_\b + \lh^\a \o_\b +
\l^\a \oh_\b + \lh^\a \oh_\b )(\Jb^I_0 + \Jb^I_1 + \Jb^I_2)({1\over 2}
Y^C Y^D \p_D \p_C U_{I\a}{}^\b + Y^C \p_C
U_{I\a}{}^\b + U_{I\a}{}^\b )].}

\listrefs

\end